\documentclass[%
reprint,
nofootinbib,
amsmath,amssymb,
aps,
pra,
]{revtex4-2}

\usepackage{graphicx}
\usepackage{amssymb,amsmath}
\usepackage[utf8]{inputenc}
\usepackage{multirow}
\usepackage{xcolor}
\usepackage{hyperref}
\hypersetup{colorlinks=true,linkcolor=blue,anchorcolor=blue,citecolor=magenta,filecolor=blue,urlcolor=blue,bookmarksnumbered=true}
\usepackage[flushleft]{threeparttable}
\usepackage{mathtools}

\usepackage{amssymb}  

\usepackage{dsfont}
\usepackage{amsthm}

\usepackage{enumitem}
\setlist[itemize]{leftmargin=10mm}

\usepackage{multirow}

\usepackage{braket}

\usepackage{mathrsfs}

\usepackage{booktabs}

\usepackage{natbib}

\usepackage[ruled,vlined]{algorithm2e}

\usepackage{longtable,array}   

\usepackage{braket}

\usepackage{enumitem}

\newcommand\startsupplement{%
	\setcounter{table}{0}\renewcommand{\thetable}{S\arabic{table}}%
	\setcounter{figure}{0}\renewcommand{\thefigure}{S\arabic{figure}}%
	\setcounter{equation}{0}\renewcommand{\theequation}{S\arabic{equation}}%
	\setcounter{section}{0}\renewcommand{\thesection}{S\arabic{section}}%
}

\begin{document}

\title{Quantum computing for gearing down complex many-fermion problems}

\author{Weijie Du\textsuperscript{1,2}}
\email[Corresponding author: ]{duweigy@gmail.com}

\author{Filippo Fornetti\textsuperscript{1}}

\author{James P. Vary\textsuperscript{2}}

\affiliation{\textsuperscript{1}Institute of Modern Physics, Chinese Academy of Sciences, Lanzhou 730000, China}
\affiliation{\textsuperscript{2}Department of Physics and Astronomy, Iowa State University, Ames, Iowa 50010, USA}

\date{\today}

\begin{abstract}
Quantum computing unlocks exciting opportunities to extend first-principles calculations of quantum many-body systems beyond classical computing limits. Despite advances in algorithms and proof-of-principle experiments, large-scale {\it ab initio} calculations on noisy intermediate-scale quantum (NISQ) devices remain challenging. We introduce a quantum-classical framework that identifies compact single-particle bases to bring otherwise intractable many-fermion problems within classical reach while preserving an accurate ground-state description. We develop an efficient adaptive search scheme driven by simulated annealing that jointly optimizes natural-orbital-guided transformations and active-space selections. A quantum-assisted ground-state estimator provides search guidance using low-depth circuits and relaxed accuracy requirements, making the framework promising for NISQ devices. We demonstrate the framework's applicability for the nuclear structure of the oxygen isotopes. Our framework opens a new route to tackling complex {\it ab initio} many-fermion problems across disciplines through quantum-assisted basis reduction.

\end{abstract}

\maketitle

{\bf Introduction.--}
{\it Ab initio} (or first-principles) methods are powerful tools for studying quantum many-body (MB) systems in quantum chemistry \cite{cook2012handbook,bauer2020quantum},  nuclear physics \cite{RevModPhys.77.427,RevModPhys.87.1067,Hjorth-Jensen:2017gss,Barrett:2013nh}, condensed matter physics \cite{fradkin2013field,michael2005computational}, and quantum field theories \cite{peskin1995introduction,BRODSKY1998299,Vary:2009gt}. Their computational demands, however, place many complex MB problems beyond the reach of classical solvers. Quantum computing offers the prospect of addressing such problems efficiently \cite{feynman1982simulating,lloyd1996universal,Jordan_2012,RevModPhys.90.015002,Cao_2019,McArdle_2020,Kreshchuk:2020dla}, but practical methods for large-scale calculations on NISQ devices \cite{Preskill2018quantumcomputingin} remain a pressing need. Here, we propose a framework to bring otherwise intractable MB problems, including those with strong nonlocal interactions, within classical reach.  Its quantum-assisted estimator is designed to operate with low-depth circuits and relaxed accuracy requirements, making the framework promising for NISQ devices.

Various quantum algorithms  have been proposed for MB structure calculations, including quantum phase estimation \cite{AspuruGuzik2005,PhysRevX.6.031007,du2021abinitionuclearstructure,PhysRevC.105.024324,PhysRevA.110.022618,yoshida2026nuclearmanybodysystemsbenchmarks}, variational quantum eigensolvers \cite{peruzzo2014variational,mcclean2016theory,PhysRevLett.120.210501,Romero_2022,Tilly_2022,PhysRevC.105.064317,Sarma_2023,P_rez_Obiol_2023}, quantum imaginary-time evolution \cite{Motta_2019,Yeter-Aydeniz:2019htv}, quantum subspace methods \cite{Cortes_2022,Epperly_2022,Kirby_2023,motta2023subspace,PhysRevA.107.032414,Du:2024zvr,Du:2024ixj}, and sample-based quantum diagonalization (SQD)  \cite{yu2025quantum,e28040367,doi:10.1126/sciadv.adu9991,Calvin2026}. However, these algorithms may struggle to solve large-scale problems on NISQ devices because, depending on the method, increasing system size can entail difficult optimization, deep quantum circuits, or substantial sampling costs.

We introduce a general-purpose quantum-classical framework to extend the reach of {\it ab initio} many-fermion calculations. It combines adaptive single-particle (SP) basis optimization with quantum-assisted ground-state (gs) estimation to gear down the SP basis dimension while retaining an accurate gs description. Quantum assistance guides the basis reduction until the resulting MB problem becomes tractable for classical solvers.

We employ natural orbitals (NOs) \cite{PhysRev.101.1730,PhysRev.97.1474,DAVIDSON1972235,RevModPhys.44.451} for adaptive basis optimization. Unlike conventional iterative NO approaches \cite{10.1021/j100880a036,Jafri1977IterativeNO,https://doi.org/10.1002/qua.560110506} with a fixed active SP dimension, our scheme couples continuous NO-guided orbital transformations with discrete selection of active SP orbitals, allowing both the size and composition of the active space to vary. This continuous--discrete search is computationally demanding for complex MB problems. To address this challenge, we develop an efficient search algorithm. Specifically, we employ inexpensive matrix operations for SP orbital updates, avoiding high-dimensional optimization of basis transformation parameters, and utilize occupation-guided stochastic moves for the selection of active orbitals, avoiding exhaustive combinatorial enumeration. We drive this coupled search with simulated annealing (SA)  \cite{Kirkpatrick1983SA}, permitting energetically unfavorable moves to facilitate escape from local minima.

For MB problems beyond classical reach, we develop a quantum-assisted estimator of gs energies and wave functions, which utilizes the Hamiltonian-encoding circuit \cite{Du:2025lhq} to sample MB configurations for subsequent classical diagonalization. Crucially, the estimator seeks only approximate information that is sufficient to guide the orbital search, rather than an accurate full-space solution. This relaxed accuracy requirement allows simpler sampling strategies with lower sampling costs. By sampling directly from the Hamiltonian-encoding circuit rather than from circuits implementing more complex Hamiltonian functions, our estimator enables the use of shallower quantum circuits, offering a promising route to noise-resilient estimation on NISQ devices. Once the active SP space becomes classically tractable, classical MB solvers take over for more accurate gs calculations and further basis optimization.

Finally, to benchmark the hybrid framework, we also develop a purely classical counterpart that operates within classically accessible MB spaces. We demonstrate both approaches through calculations of oxygen isotopes with realistic nuclear interactions. Our framework provides a new route toward extending {\it ab initio} calculations to larger, strongly interacting many-fermion systems across multiple disciplines utilizing NISQ devices.

{\bf Theory.--}
We consider a system of $\eta$ fermions in, e.g., quantum chemistry \cite{cook2012handbook,bauer2020quantum}, nuclear physics \cite{Hjorth-Jensen:2017gss,Barrett:2013nh}, and hadron physics \cite{Vary:2009gt,BRODSKY1998299}. In an original SP basis $\{\ket{p}\}$, the particle-number-conserving Hamiltonian up to two-body terms is
\begin{equation}
	H=\sum_{p,q}\langle p|T|q\rangle a_p^\dagger a_q
	+\sum_{p<q}\sum_{r<s}\langle pq|V|rs\rangle
	a_p^\dagger a_q^\dagger a_s a_r,
	\label{eq:HamiltonianLabeling}
\end{equation}
with $p,q,r,s\in\Omega=\{0,1,\cdots,N_{\rm sp}-1\}$. 
$\langle p|T|q\rangle$ and $\langle pq|V|rs\rangle$ are the one-body and antisymmetrized two-body matrix elements, respectively. The fermion operators $a_p^\dagger$ and $a_q$ obey the canonical anticommutation relations. We retain only one- and two-body terms for the simplicity of discussion. 

A real orthogonal transformation $U\in O(N_{\rm sp})$ defines the working SP basis $\{ \ket{\alpha} \}$ with $\alpha\in\Omega$ through $\ket{\alpha}=\sum_p U_{p\alpha}\ket{p}$. The corresponding operators, $c_\alpha^\dagger=\sum_p U_{p\alpha}a_p^\dagger$ and $c_\beta=\sum_p U_{p\beta}a_p$, preserve the anticommutation relations. In this basis, the Hamiltonian reads
\begin{equation}
	H(U, \Omega)=\sum_{\alpha,\beta}\widetilde{T}_{\alpha\beta}
	c_\alpha^\dagger c_\beta
	+\sum_{\alpha<\beta}\sum_{\gamma<\delta}
	\widetilde{V}_{\alpha\beta\gamma\delta}
	c_\alpha^\dagger c_\beta^\dagger c_\delta c_\gamma,
	\label{eq:transformed_Hamiltonian}
\end{equation}
where $\widetilde{T}_{\alpha\beta}$ and
$\widetilde{V}_{\alpha\beta\gamma\delta}$ are the transformed one- and two-body matrix elements. The transformation leaves the full MB spectrum unchanged.

We partition the working SP basis into active and inactive orbitals, indexed by the active index set (AIS) $A\subseteq\Omega$ and its complementary inactive index set $F=\Omega\setminus A$, respectively. The active MB space consists of configurations constructed exclusively from the $k=|A|$ active SP orbitals, reducing the MB dimension from $\binom{N_{\rm sp}}{\eta}$ to $\binom{k}{\eta}$ before imposing symmetry restrictions. The Hamiltonian restricted to this space is
\begin{equation}
	H(U,A)=\mathcal P(U,A)H(U,\Omega)\mathcal P(U,A),
\end{equation}
where $\mathcal P(U,A)$ projects onto the active MB space; other operators are projected similarly. We denote its gs energy and vector by $E_0(U,A)$ and $\ket{\Psi_0(U,A)}$, respectively. By the variational principle, we have $E_0(U,A)\geq E_{\rm exact}$, where $E_{\rm exact}$ is the exact gs energy of the full-space $H$.

Our goal is to identify the smallest active SP space satisfying a prescribed energy tolerance $\xi$ by jointly optimizing over the continuous variable of orbital transformation $U\in O(N_{\rm sp})$ and the discrete variable of AIS $A\subseteq\Omega$, with $|A|\geq\eta$:
\begin{equation}
	\min_{U,A}|A|\quad\text{subject to}\quad
	E_0(U,A)-E_{\rm exact}<\xi .
	\label{eq:goal}
\end{equation}

In practice, we use NOs to guide this coupled continuous--discrete optimization, since they provide a compact SP basis for expanding correlated MB states. Starting from the complete SP basis of dimension $k_{\rm tot}=N_{\rm sp}$, we progressively reduce the active SP space toward a target dimension $k_{\rm tar}<k_{\rm crit}$, where $k_{\rm crit}$ is the largest active SP dimension that keeps the $\eta$-fermion MB space tractable for the classical solver. 
In the working basis $\{\ket{\alpha}\}$, the NOs are obtained by diagonalizing the one-body reduced density matrix (1-RDM)
\begin{equation}
	\gamma_{\alpha\beta}(U,A)
	=\bra{\Psi_0(U,A)}c_\alpha^\dagger c_\beta\ket{\Psi_0(U,A)}.
\end{equation}

For the real-valued states considered here, we obtain the NOs in the original SP basis by diagonalizing
\begin{equation}
	V^\dagger\bigl[U\gamma(U,A)U^\dagger\bigr]V=\Gamma,
	\qquad \Gamma=\operatorname{diag}(n_0,\ldots,n_{N_{\rm sp}-1}).
\end{equation}
The columns of $V$ give the NOs, while the eigenvalues $n_\alpha\in[0,1]$ are the natural occupation numbers, with $\sum_\alpha n_\alpha=\eta$. 
These occupations characterize the particle distribution among the NOs and guide the selection of active SP orbitals.

For large-scale MB problems, however, obtaining an accurate full-space gs vector can be computationally challenging, making the construction of an accurate full-space 1-RDM impractical. We instead develop an iterative approach to identify a compact, energetically favorable active SP basis through NO-guided orbital transformations and AIS selections. We drive this continuous--discrete search with SA, where we also develop two complementary schemes for efficient basis optimization.

We start the iteration from the original SP basis with $U_0=I$ and $A_0=\Omega$. We first obtain an approximate reference 1-RDM $\gamma_{\rm ref}$ from a full-space calculation using the quantum-assisted estimator (introduced below) for the classically intractable full-space problem. This reference provides auxiliary occupation information to guide active SP-basis selection. In this work, we keep $\gamma_{\rm ref}$ fixed, although it could be updated during the SA-guided search.

We denote the current configuration at step $n$ by $(U_n,A_n)$, where $U_n$ specifies the working SP basis and $A_n$ identifies the active SP bases used in the MB calculation. The iteration proceeds as
\begin{equation}
	(U_{n+1},A_{n+1})=\mathcal{T}_{t_n}(U_n,A_n),
	\label{eq}
\end{equation}
where $\mathcal{T}_{t_n}$ combines the correlated MB calculation, 1-RDM construction, NO transformation, and active-space update with a stochastic acceptance test at temperature $t_n$. Each iteration consists of the following steps.

\paragraph{\it AIS update.}
At step $n$, the current MB calculation projected onto the active SP space yields a 1-RDM $\gamma_n=\gamma(U_n,A_n)$, where the diagonal elements $q_\alpha=(\gamma_n)_{\alpha\alpha}$ give the occupations of active SP orbitals $ \{ \ket{\alpha} \}$ with  $\alpha\in A_n$. Since the MB calculation excludes the inactive SP orbitals $ \{ \ket{\beta} \}$ with  $\beta \in F_n$, their occupations vanish in the resulting state, and hence $(\gamma_n)_{\beta\beta}=0$ for $\beta\in F_n$. We estimate the occupations of the inactive orbitals using $ q_\beta^{\rm ref} = (U_n^\dagger\gamma_{\rm ref}U_n)_{\beta\beta} $. 

To propose a new AIS $A_n'$, we introduce three probabilistic move types to update the current AIS $A_n$. First, ``{\it Add}'' selects an inactive orbital $\ket{\beta}$ with $\beta\in F_n$ with weight proportional to its reference occupation entropy \cite{Comment1}
\begin{equation}
	S_\beta=-q_\beta^{\rm ref}\ln q_\beta^{\rm ref}	-(1-q_\beta^{\rm ref})\ln(1-q_\beta^{\rm ref}) .
\end{equation}
We note that $S_\beta $ favors fractionally occupied orbitals: it is largest at $1/2$ and vanishes at zero and one. Second, ``{\it Remove}'' selects an active orbital indexed by $\alpha\in A_n$ with weight proportional to $\exp[-\kappa q_\alpha/t_n]$, where $\kappa>0$ controls the preference for weakly occupied orbitals. Third, ``{\it Swap}'' replaces an active orbital with an inactive one using the corresponding {\it Remove} and {\it Add} weights. 

These moves guide the active SP space toward a target dimension $k_{\rm tar}$. Compared with the direct search based on exhaustive enumeration of active-orbital subsets, our adaptive search avoids the combinatorial scan of possible AIS configurations, thereby improving search efficiency.

\paragraph{\it Working basis update.}
With the updated AIS $A_n'$ of dimension $k'=|A_n'|$, we construct and solve the MB Hamiltonian $H(U_n,A_n')$ to obtain its gs vector $\ket{\Psi_0(U_n,A_n')}$ and the corresponding 1-RDM $\gamma_n'$. We transform $\gamma_n'$ into the original SP basis and diagonalize it to obtain the NO matrix $V_n$, which provides the current NO direction for updating $U_n$. To avoid arbitrary eigenvector reorderings and sign changes during calculations, we match the columns and signs of $V_n$ to those of $U_n$ using orbital overlaps.

A conventional approach to updating the working basis $U_n$ toward the matched NOs represented by $V_n$ is to parameterize an orthogonal orbital rotation as $W=\exp(Y)$, where the real antisymmetric generator $Y$ contains $N_{\rm sp}(N_{\rm sp}-1)/2$ independent parameters \cite{PhysRevLett.134.206401}. Since optimizing these parameters typically requires repeated evaluations of the matrix exponential, this approach can be computationally costly within the iterative SA search. We therefore propose an efficient scheme to construct the trial working basis $U_n'$ without explicitly optimizing the rotation parameters.

Specifically, we keep the inactive orbitals indexed by $F_n'=\Omega\setminus A_n'$ fixed and partition $U_n$ and $V_n$ into the column blocks $U_{n;A_n'} = U_n(:,A_n'),\  U_{n;F_n'} =U_n(:,F_n'), \ V_{n;A_n'} =V_n(:,A_n') $, where $U_{n;A_n'}$ contains the current active orbitals and $V_{n;A_n'}$ contains their matched NOs. In what follows, we use the matched NOs as target orbitals to update the active orbitals while keeping the inactive ones fixed.

We first interpolate the active orbitals toward their matched NOs in the ambient matrix space
\begin{equation}
	\widetilde{U}_{n;A_n'}	=	(1-\alpha_n)U_{n;A_n'}	+\alpha_n V_{n;A_n'},
	\qquad 
	0\leq\alpha_n\leq1,
	\label{eq:ambient_interpolation}
\end{equation}
where the SA temperature controls the update amplitude $\alpha_n$. 
Since the interpolated columns may have components along the fixed inactive orbitals, we then project them onto the orthogonal complement of the inactive space
\begin{equation}
	\widehat{U}_{n;A_n'}	=	(\mathbb{I}-P_{F_n'})	\widetilde{U}_{n;A_n'},	
	\qquad	
	P_{F_n'}	=	U_{n;F_n'}U_{n;F_n'}^\dagger.
	\label{eq:projectionToA}
\end{equation}

Next, we orthonormalize the projected columns using polar decomposition \cite{Higham1986Polar,Schonemann1966Procrustes}
\begin{equation}
	U'_{n;A_n'} 	= 	\widehat{U}_{n;A_n'}	\left(	\widehat{U}_{n;A_n'}^\dagger	\widehat{U}_{n;A_n'}	\right)^{-1/2}  .
	\label{eq:polar}
\end{equation}

Finally, we assemble the trial working basis $U_n'$ by setting $U_n'(:,A_n')=U'_{n;A_n'}$ and $U_n'(:,F_n')=U_{n;F_n'}$. This low-cost working-basis update preserves the active-subspace span while steering the active orbitals toward their matched NOs, facilitating an efficient search for a compact SP basis during SA optimization.

\paragraph{\it Metropolis acceptance.}
For the proposal $(U_n',A_n')$, we construct and solve the MB Hamiltonian $H(U_n',A_n')$ to obtain its ground-state energy $E_0(U_n',A_n')$, with which we evaluate
\begin{equation}
	\Delta E
	=
	E_0(U_n',A_n')-E_0(U_n,A_n).
\end{equation}
The proposal $(U_n',A_n')$ is accepted with the Metropolis probability
$P_{\rm acc}=\min(1,e^{-\Delta E/t_n})$.
If accepted, we set
$(U_{n+1},A_{n+1})=(U_n',A_n')$;
otherwise, we retain
$(U_{n+1},A_{n+1})=(U_n,A_n)$.

The annealing temperature decreases according to $t_{n+1}=\zeta t_n$, with $\zeta\in(0,1)$, and reheating is applied when the search stagnates. At finite temperature, uphill proposals can be accepted, helping the search escape local minima. As $t_n$ decreases, their acceptance becomes less likely. The temperature also controls $\alpha_n$ and its stochastic perturbation, enabling larger orbital updates at high temperature and finer adjustments at low temperature.

{\bf Hybrid gs estimator.--}
We propose a quantum-assisted eigensolver to estimate the gs energy and vector of $\mathcal{H}= H(U_n,A_n)$ for $|A_n|>k_{\rm crit}$.  The estimator combines our Hamiltonian input scheme \cite{Du:2025lhq,Du:2023bpw} with configuration sampling and classical configuration interaction (CI) calculations.

In this estimator, we encode the fermion operators using Boolean masking \cite{helgaker2014molecular} and construct a quantum-walk-based block encoding of the Hamiltonian as [Eq. (8), \citenum{Du:2025lhq}]
\begin{equation}
	(\bra{\mathcal{G}}_s \otimes \bra{0}_a ) \mathcal{T}_b^{\dagger} \mathcal{T}_f ( \ket{\mathcal{F}}_s \otimes \ket{0}_a ) = \bra{\mathcal{G}} \mathcal{H} \ket{\mathcal{F}} /\mathcal{C} ,
	\label{eq:BEofHamiltonian}
\end{equation}
where $\ket{\mathcal{F}}$ and $\ket{\mathcal{G}}$ are MB states, and $\mathcal{T}_b$ and $\mathcal{T}_f$ are the isometries. The subscripts $s$ and $a$ label the system and ancillary registers, respectively. The normalization constant $\mathcal{C}$ depends on the number and maximum magnitude of the few-body Hamiltonian matrix elements. For convenience, we define $\mathcal{H}'=\mathcal{H}/\mathcal{C}$.

We aim to identify a tractable CI space that captures useful gs information to guide the search. To this end, we consider gs preparation through imaginary-time filtering, $\exp[-\tau\mathcal{H}']\ket{\Phi_{\rm pivot}}$, using our Hamiltonian input scheme, where $\tau>0$ and $\ket{\Phi_{\rm pivot}}$ is a pivot state with nonzero overlap with the gs of $\mathcal{H}'$. Since $\exp[-\tau\mathcal{H}']$ admits an expansion in Chebyshev polynomials $\{T_n(\mathcal{H}')\}$ \cite{kenken2013mathematical,PRXQuantum.2.040203},  the vectors $\{T_n(\mathcal{H}')\ket{\Phi_{\rm pivot}} \}$ span a Krylov subspace that can approximately represent the gs vector.

Rather than implementing the full imaginary-time filtering, we seek only a representative subset of important MB configurations that capture the dominant support for the gs vector. Specifically, we sample $\mathcal{H}'\ket{\phi_{\rm tr}}$ (with $T_1(\mathcal{H}')=\mathcal{H}'$) for a set of easily preparable trial pivots $\{\ket{\phi_{\rm tr}}\}$ on quantum computers and retain those MB configurations with measurement probabilities exceeding a moderate threshold $\varepsilon$. For each trial, we perform a CI calculation in the space spanned by the retained configurations to obtain the trial gs energy $E_{\rm tr}$ and vector $\ket{\Psi_{\rm tr}}$. We then select the lowest trial energy $E_{\rm tr,min}$ and its corresponding state $\ket{\Psi_{\rm tr,min}}$ for our estimates of the gs energy and vector of $\mathcal{H}$.

Our estimator requires only the Hamiltonian-input circuits, thereby avoiding deep circuits for complex Hamiltonian functions and enabling shallow-circuit implementations.
As it operates with relaxed accuracy requirements, it avoids using elaborate sampling schemes or exhaustive configuration sampling. These features may reduce the effects of sampling errors and hardware noise, making the approach promising for NISQ devices \cite{Preskill2018quantumcomputingin}.

\begin{figure}[ht]
	\centering
	\includegraphics*[width=0.95\linewidth]{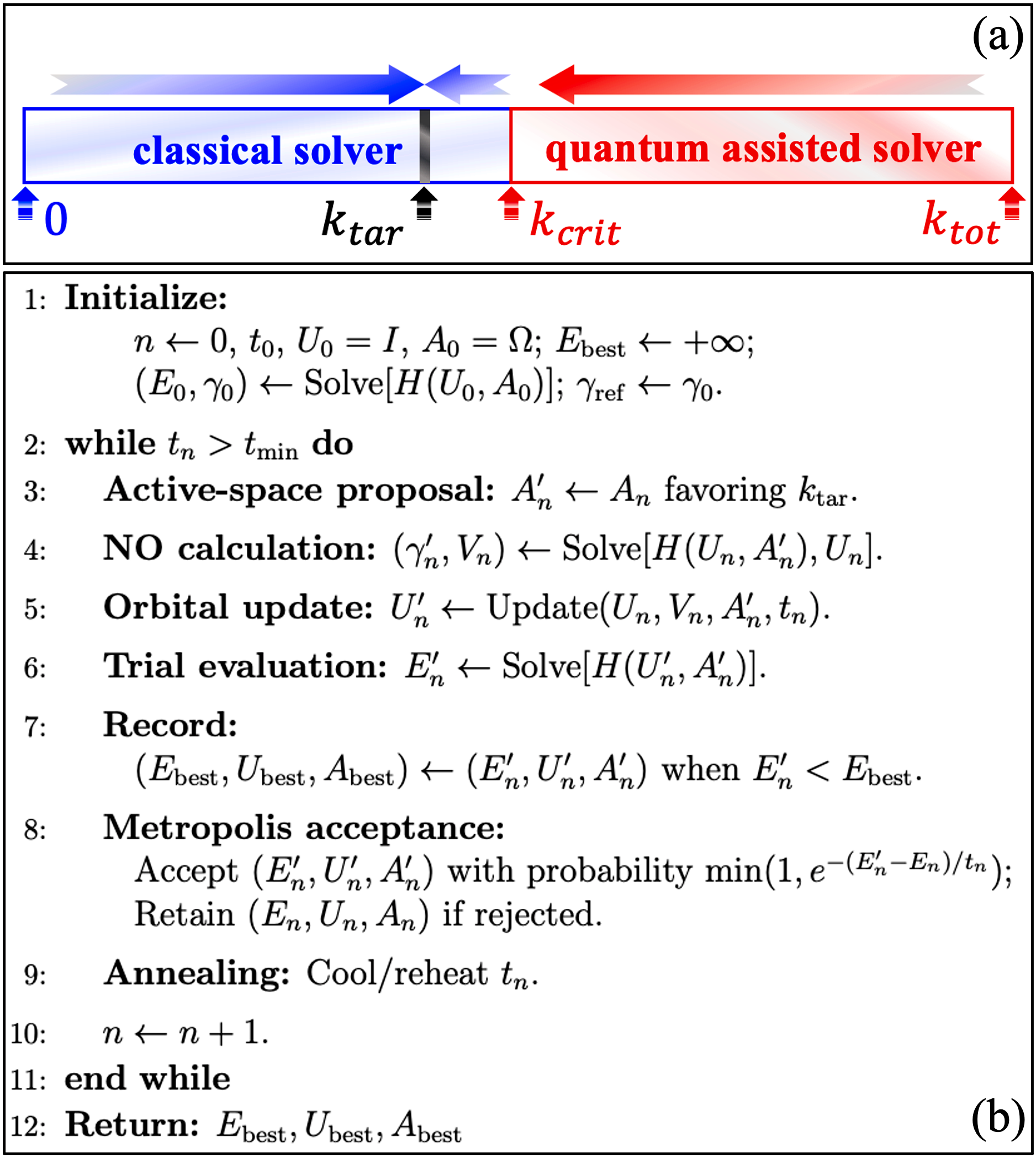}
	\caption{
		(a) Schematic workflow of the hybrid framework; (b) Pseudocode of the hybrid top-down scheme.
	}
	\label{fig:scheme_and_pseudocode}
\end{figure}

{\bf Hybrid framework.--}
Combining the NO-guided SA search with the quantum-assisted gs estimator, we propose a hybrid quantum-classical framework to estimate the gs energy and vector of many-fermion systems beyond the reach of classical solvers. The workflow and pseudocode are summarized in Fig. \ref{fig:scheme_and_pseudocode}(a, b).

Starting from the original SP basis and second-quantized Hamiltonian ($U_0=I$, $A_0=\Omega$), the top-down search reduces the active SP dimension from $k_{\rm tot}=N_{\rm sp}$ toward $k_{\rm tar}$, with $\eta\leq k_{\rm tar}<k_{\rm crit}$. For $|A_n|>k_{\rm crit}$, the quantum-assisted estimator provides approximate gs energies and vectors of the $\eta$-fermion system to guide the search; classical Full Configuration Interaction (FCI) takes over for $|A_n|\leq k_{\rm crit}$. Throughout the search, we retain the lowest-energy configuration satisfying $\eta\leq |A|\leq k_{\rm tar}$. The classical MB calculation in the selected compact SP basis yields the gs energy and vector, from which other gs observables can be evaluated.

For comparison, we also propose a purely classical ``bottom-up" counterpart, which starts from $U_0=I$ and $A_0\subset\Omega$, with $\eta\leq|A_0|<k_{\rm tar}$. All MB calculations remain classically tractable as the active-SP-space dimension approaches $k_{\rm tar}$. This approach does not require an initial full-space reference calculation. Without prior occupation information, {\it Add} and {\it Swap} moves sample inactive orbitals uniformly. Such less-directed sampling can lead to inefficient exploration, motivating further improvements to the bottom-up search strategy \cite{Comment1}.

\begin{figure*}[ht]
	\centering
	\includegraphics[width=0.49\textwidth]{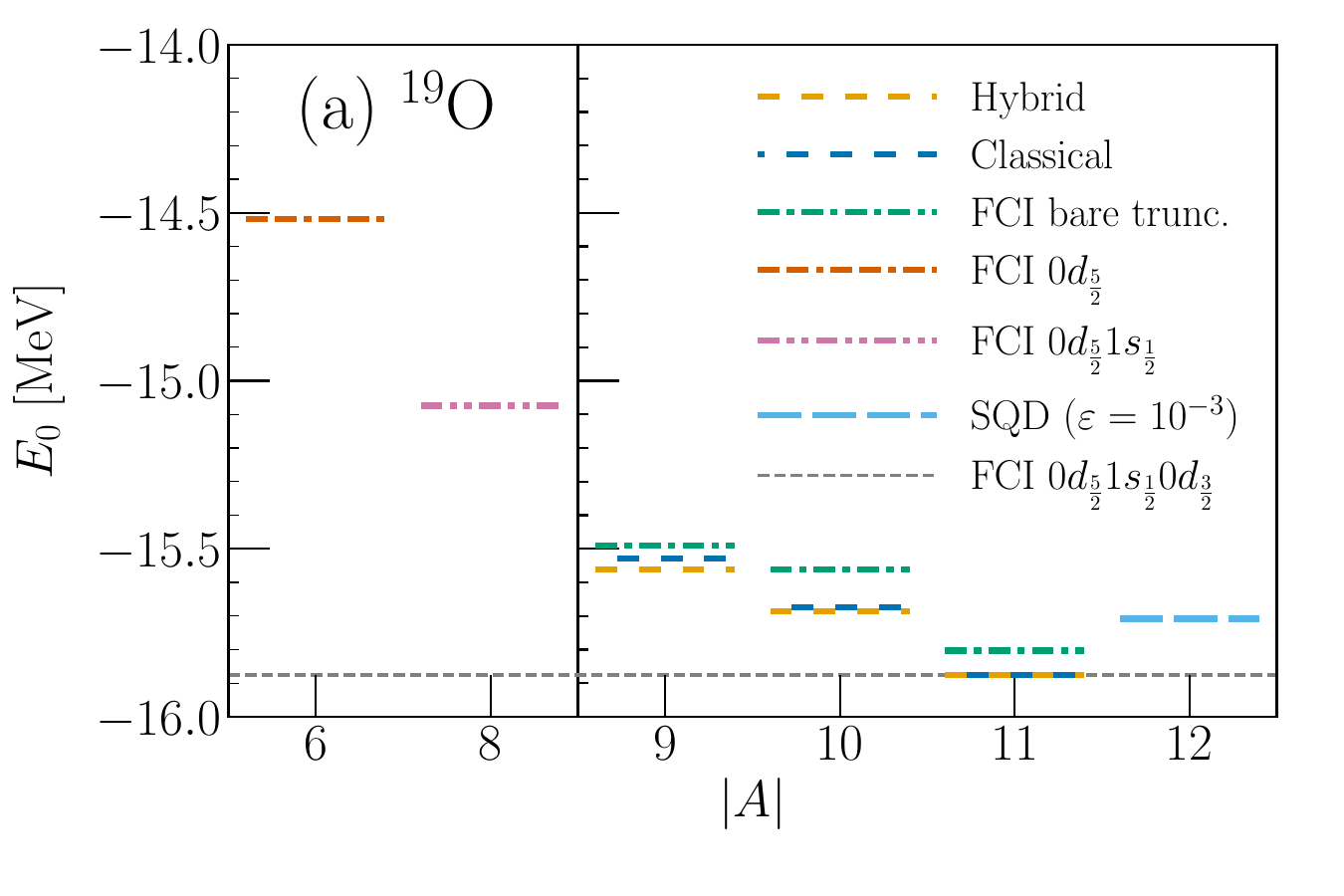}
	\hspace{-0.035\textwidth}
	\includegraphics[width=0.49\textwidth]{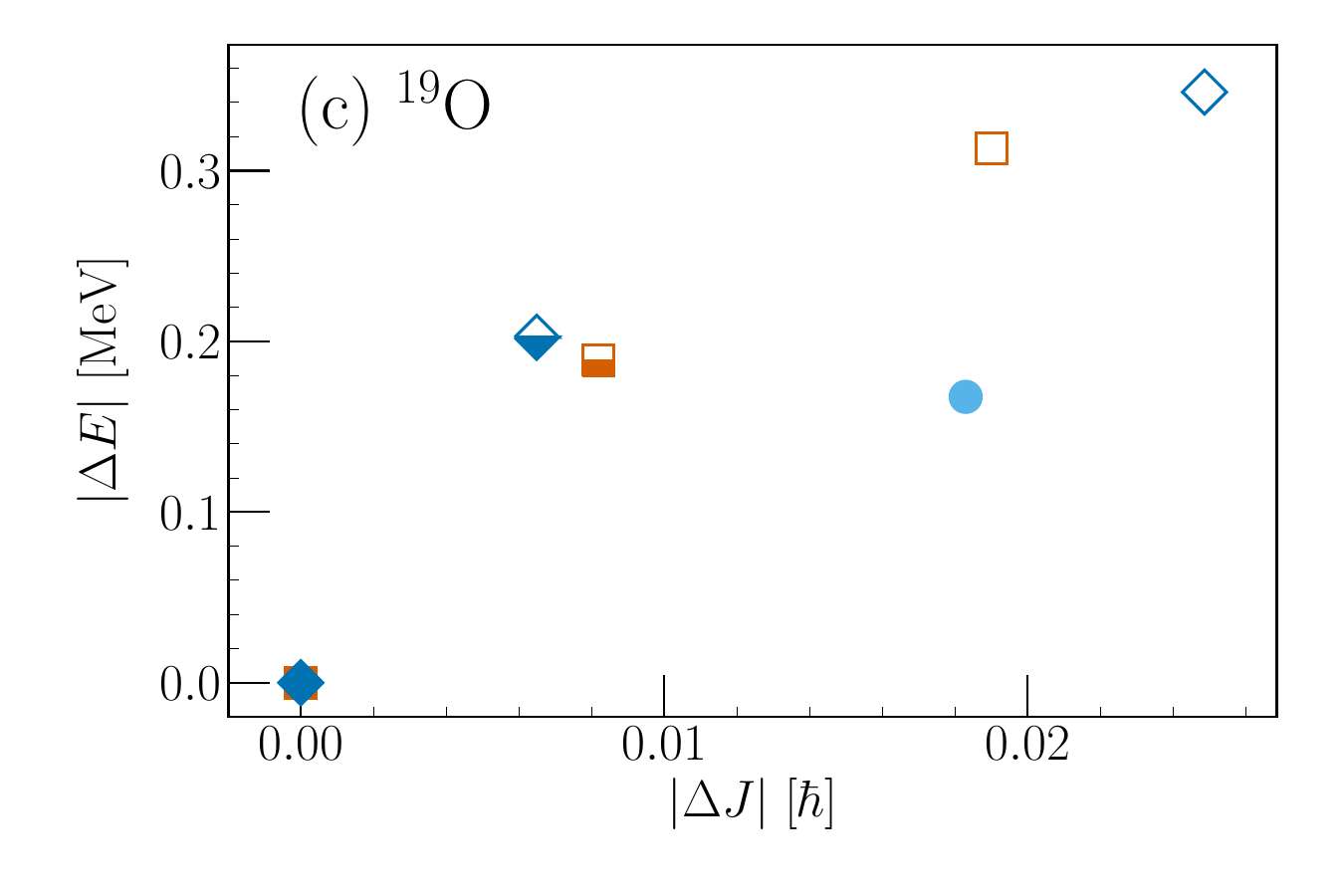}
	
	\vspace{-0.5cm}
	
	\includegraphics[width=0.49\textwidth]{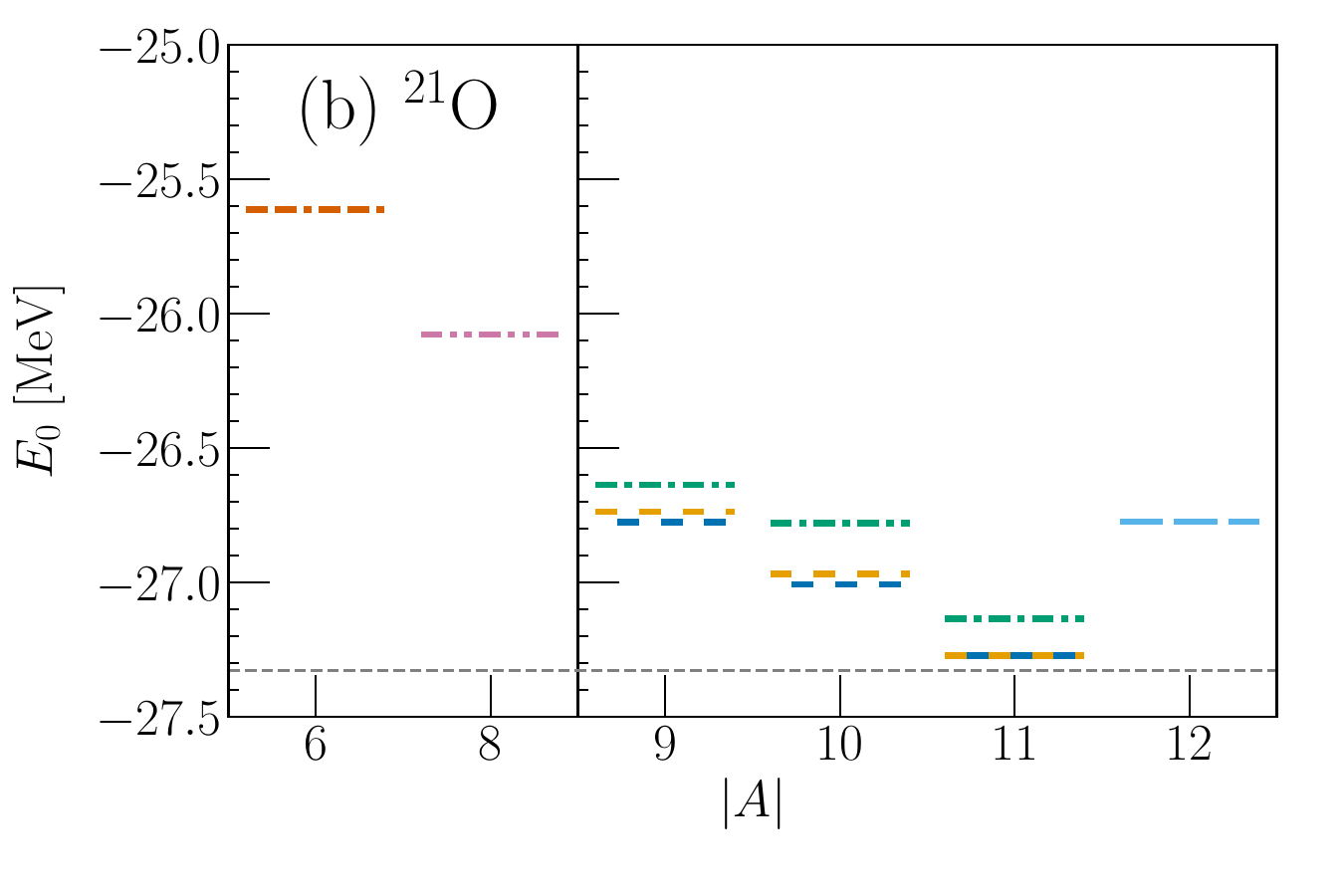}
	\hspace{-0.035\textwidth}
	\includegraphics[width=0.49\textwidth]{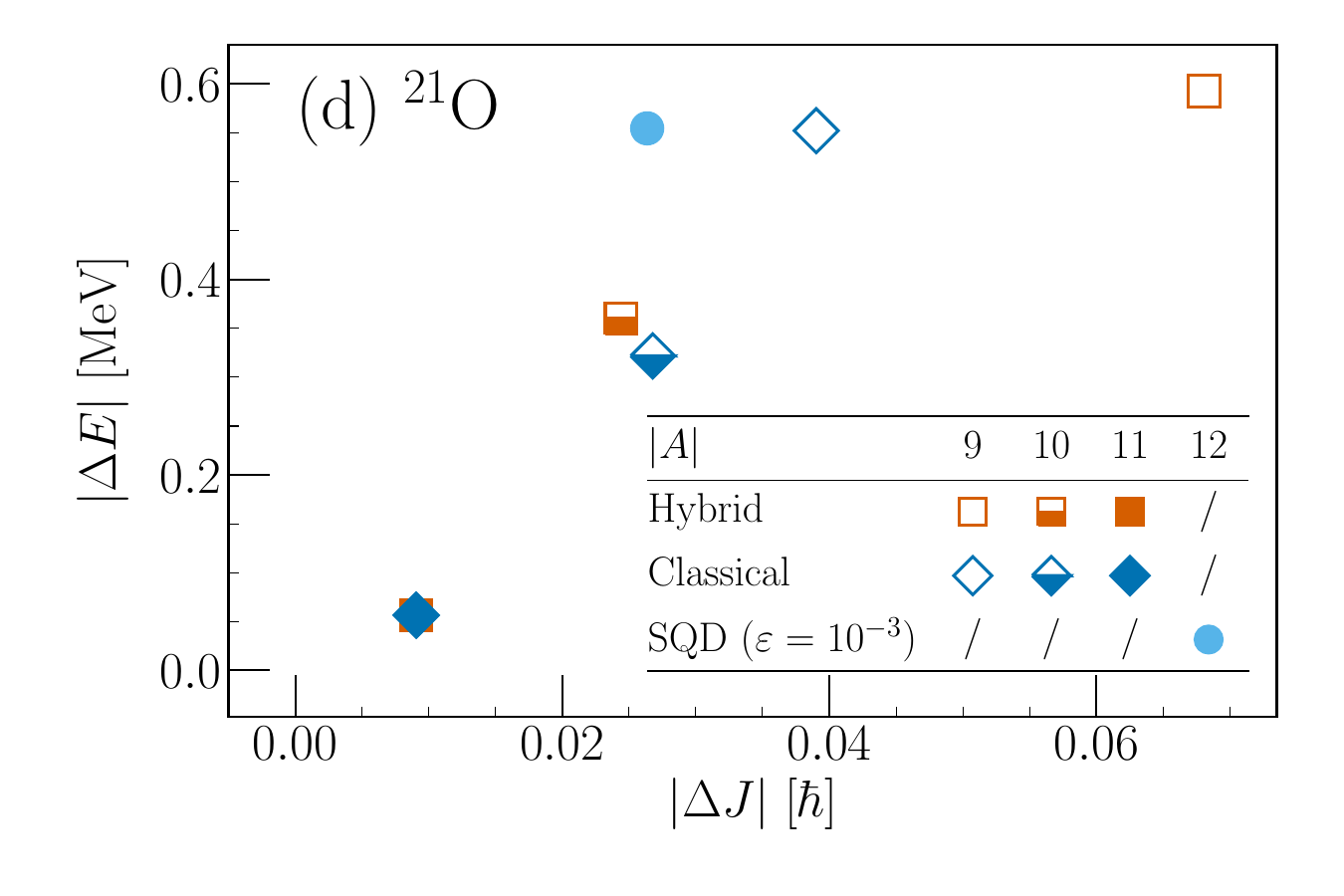}
	
	\caption{
		(a, b) The gs energies obtained via different methods and inputs as a function of $|A|$ for $^{19}$O and $^{21}$O. (c, d) The absolute deviations of the gs energies {($|\Delta E|$) and $J$ values ($|\Delta J|$)} obtained with the hybrid and classical approaches from their respective exact values for $^{19}$O and $^{21}$O. 
	}
	\label{fig:results}
\end{figure*}

{\bf Numerical demonstration.--}
We illustrate the framework through calculations of $^{19}$O and $^{21}$O as valence neutron systems in the $sd$ space outside an ${}^{16}$O core.
We use the three-dimensional harmonic-oscillator SP basis \cite{suhonen2007nucleons}, retaining the $0d_{5/2}$, $1s_{1/2}$, and $0d_{3/2}$ orbitals with $k_{\rm tot} = 12$. The interaction is taken from Ref. \cite{Shin:2024zpe} and transformed into this SP basis representation. The resulting 632 few-body matrix elements are listed in Table \ref{tab:few_body_matrix_elements} in the Supplementary Material.

Figure \ref{fig:results}(a-b) shows the gs energies obtained with the hybrid top-down approach for $k_{\rm crit}=11$ and active-space dimensions $|A|=9$, 10, and 11. The quantum-assisted estimator is simulated using Qiskit \cite{Qiskit} in noiseless mode, and its results are cross-checked against classical calculations. We also provide the results from the purely classical bottom-up approach.

For comparison, we provide results via other methods and inputs. Specifically, we present a set of FCI calculations retaining the SP bases from the $0d_{5/2}$ shell (with $|A|=6$), the $0d_{5/2} 1s_{1/2}$ shell (with $|A|=8$), as well as the complete $sd$ valence shell $0d_{5/2} 1s_{1/2} 0d_{3/2}$ (with $|A|=12$) that present the exact gs energies. Meanwhile, we present the  best gs energies obtained with our quantum-assisted eigensolver, using the SP bases of the full $sd$ valence shell  (with $|A|=12$) and adopting the sampling threshold $\varepsilon$ such that only those MB bases with measurement probabilities exceeding $10^{-3}$ are retained for subsequent CI calculations. In the middle of the shell structure evolution (i.e., $|A| =9, 10,$ and $11$), we present the lowest energies obtained from ``bare" truncated Hamiltonians that are based on the MB states formed by a fixed number $|A|$ of SP bases in the $sd$ shell (note that there are $\binom{12}{|A|}$ different combinations for each chosen $|A|$).

We find that the gs energies obtained with our hybrid and classical frameworks exhibit a variational trend with $|A|$. We also note small differences in the gs energies between the two approaches, which are expected to arise from their distinct stochastic trajectories in basis-space exploration and optimization strategies.

By accounting for correlations among the SP bases, our calculations using the NO bases yield lower gs energies than the best results obtained via the bare-truncation approach. Notably, our results at $|A|=11$ closely reproduce the corresponding exact gs energies from full-space calculations at $|A|=12$. The calculated gs $J$ values at $|A|=11$ also agree well with the corresponding exact values  [Fig. \ref{fig:results}(c-d)], indicating the quality of our MB wave functions for computing other structure observables.

Although the demonstration is limited to a restricted effective shell-model problem with $N_{\rm sp}=12$, we expect our algorithms to be applicable to larger MB systems with $\eta \ll N_{\rm sp}$, where full-space FCI problems can be reduced to restricted-scale calculations accessible to classical computers.

{\bf Conclusion and outlook.--}
In sum, we introduce a quantum-classical framework that identifies a compact SP basis for {\it ab initio} many-fermion calculations, bringing them within classical reach while retaining an accurate gs description. This framework combines an efficient SA-driven search over continuous NO-guided orbital transformations and discrete active-SP-space selections of varying dimension with a quantum-assisted gs estimator. The estimator is designed for low-depth circuit implementation and operates with relaxed accuracy requirements, making the framework promising for NISQ devices. Calculations of oxygen isotopes using realistic nuclear interactions demonstrate that the framework produces accurate gs energies and total angular momenta with fewer SP orbitals.

Future applications include various many-electron and many-nucleon systems, as well as hadron structure within light-front Hamiltonian approaches \cite{BRODSKY1998299,Vary:2009gt}. Further developments will exploit machine learning \cite{Goodfellow2016DeepLearning,Prince2023UnderstandingDL} for efficient Hamiltonian sampling \cite{Kiss_2023,Campbell_2019} and orbital optimization for NISQ implementations.

{\bf Acknowledgments.--}
We are grateful for fruitful discussions with Prof. A. M. Shirokov.
WD was supported by the startup funding from the Chinese Academy of Sciences. 
This work was supported in part by the U.S. DOE Grant NO. DE-SC026930 under the Office of Nuclear Physics Quantum Horizons program for the “Nuclei and Hadrons with Quantum Computers (NuHaQ)” project. This project was also supported in part by NSF Grant No. 2435255 (NQVL-QSTD: Q-BLUE).

\bibliography{apssamp}

\begin{thebibliography}{70}
\bibitem [{Com()}]{Comment1}%
  \BibitemOpen
  \href@noop {} {}\bibinfo {note} {The approach can also be generalized to an
  initial active set $A_0\subset\Omega$. Without reference occupation
  information, candidate orbitals for Add moves are sampled
  uniformly.}\BibitemShut {Stop}%
\end{thebibliography}

\newpage

\onecolumngrid

\appendix

\clearpage
\startsupplement
\section{Supplementary Material: effective $sd$-shell interaction}

We compute the eigenenergies and the corresponding total angular momenta of the ground states of $ ^{19}$O and $ ^{21}$O in this work. We adopt a subspace of the $sd$-valence shell. In particular, we retain the single-particle (SP) basis states associated with the $0d_{5/2}1s_{1/2}0d_{3/2}$ orbitals. The adopted SP basis states are listed in Table \ref{tab:SP_basis}.

\begin{table}[ht]
	\centering
	\caption{The SP basis states in the $0d_{5/2}1s_{1/2}0d_{3/2}$ valence shell. The quantum numbers of each SP state are presented, while the spin and isospin of the neutrons are understood to be $1/2$, and the isospin projection is $-1/2$. Each SP basis state is indexed and mapped to a specific qubit.}
	\begin{tabular}{cccccccc} 
		\hline \hline
		& SP basis (qubit) & $n$ & $l$ & $2j$ & $2m$ &  \\ \hline
		\multirow{2}{*}{$1s_{1/2}$} &  0              & 1   & 0   & 1    & $-1$    \\
		& 1                & 1   & 0   & 1    & $+1$    \\ \hline
		\multirow{4}{*}{$0d_{3/2}$} &  2                & 0   & 2   & 3   & $-3$    \\
		& 3               & 0   & 2   & 3    & $-1$    \\
		& 4               & 0   & 2   & 3    & $+1$    \\
		& 5               & 0   & 2   & 3    & $+3$    \\ \hline		
		\multirow{6}{*}{$0d_{5/2}$} & 6                & 0   & 2   & 5    & $-5$    \\
		& 7               & 0   & 2   & 5    & $-3$    \\
		& 8               & 0   & 2   & 5    & $-1$    \\
		& 9              & 0   & 2   & 5    & $+1$    \\
		& 10              & 0   & 2   & 5    & $+3$    \\
		& 11               & 0   & 2   & 5    & $+5$    \\ 
		\hline \hline
	\end{tabular} 
	\label{tab:SP_basis}
\end{table}

The Hamiltonian of the many-neutron system is 
\begin{equation}
	H = E_{\rm core} + \sum _{i=0} ^{11} \varepsilon _i a^{\dagger}_i a_i + \sum _{p<q} ^{11} \sum _{r<s} ^{11} \langle p q | V | r s \rangle a^{\dagger}_p a^{\dagger}_q a _s a _r ,
	\label{eq:hamiltonian_expression_SM}
\end{equation}
where we adopt the phenomenologically successful choices from Ref. \cite{Shin:2024zpe} and set the SP energies $\varepsilon _i = -3.2079 $ MeV for $i \in [0, 1]$, $\varepsilon _i = + 2.1117$ MeV for $i \in [2, 5]$, and $\varepsilon _i = -3.9257$ MeV for $i \in [6, 11]$. We set the core energy to be $E_{\rm core}=0$ in our work. The two-body interaction matrix elements are derived from the effective $sd$-shell nucleon-nucleon (NN) interaction \cite{Shin:2024zpe}. In Table \ref{tab:sd_nn_effective_interaction}, we list these interaction matrix elements in the coupled-$JT$ basis. 

\begin{table}[ht]
	\centering
	\caption{Effective $sd$-shell NN interaction matrix elements (in MeV) in the coupled-$JT$ basis.These elements are taken from Table S1 of Ref. \cite{Shin:2024zpe} for $J= 0,\ 1,\ 2,\ 3, \ 4$ and $T=1$.}
	\begin{tabular}{cccccc|cccccc|cccccc}
		\hline \hline 
		$2j_a$ & $2j_b$ & $2j_c$ & $2j_d$ & $J$ & $V$     & $2j_a$ & $2j_b$ & $2j_c$ & $2j_d$ & $J$ & $V$     & $2j_a$ & $2j_b$ & $2j_c$ & $2j_d$ & $J$ & $V$     \\ \hline \hline 
		1      & 1      & 1      & 1      & 0   & -1.9100  & 1      & 3      & 3      & 3      & 2   & -0.013000 & 3      & 3      & 5      & 5      & 2   & -0.55100 \\
		1      & 1      & 3      & 3      & 0   & -0.27400 & 3      & 3      & 1      & 3      & 2   & -0.013000 & 5      & 5      & 3      & 3      & 2   & -0.55100 \\
		3      & 3      & 1      & 1      & 0   & -0.27400 & 1      & 3      & 3      & 5      & 2   & 0.47700  & 3      & 5      & 3      & 5      & 2   & 0.11300  \\
		1      & 1      & 5      & 5      & 0   & -1.5640  & 3      & 5      & 1      & 3      & 2   & 0.47700  & 3      & 5      & 5      & 5      & 2   & 0.26900  \\
		5      & 5      & 1      & 1      & 0   & -1.5640  & 1      & 3      & 5      & 5      & 2   & -1.2110  & 5      & 5      & 3      & 5      & 2   & 0.26900  \\
		3      & 3      & 3      & 3      & 0   & -1.6540  & 5      & 5      & 1      & 3      & 2   & -1.2110  & 5      & 5      & 5      & 5      & 2   & -1.1190  \\
		3      & 3      & 5      & 5      & 0   & -3.2790  & 1      & 5      & 1      & 5      & 2   & -1.0320  & 1      & 5      & 1      & 5      & 3   & 0.75100  \\
		5      & 5      & 3      & 3      & 0   & -3.2790  & 1      & 5      & 3      & 3      & 2   & -1.1140  & 1      & 5      & 3      & 5      & 3   & 0.29800  \\
		5      & 5      & 5      & 5      & 0   & -2.4390  & 3      & 3      & 1      & 5      & 2   & -1.1140  & 3      & 5      & 1      & 5      & 3   & 0.29800  \\
		1      & 3      & 1      & 3      & 1   & 0.36800  & 1      & 5      & 3      & 5      & 2   & 0.38300  & 3      & 5      & 3      & 5      & 3   & 0.52700  \\
		1      & 3      & 3      & 5      & 1   & -0.081000 & 3      & 5      & 1      & 5      & 2   & 0.38300  & 3      & 5      & 3      & 5      & 4   & -1.3090  \\
		3      & 5      & 1      & 3      & 1   & -0.081000 & 1      & 5      & 5      & 5      & 2   & -0.54400 & 3      & 5      & 5      & 5      & 4   & 1.3800   \\
		3      & 5      & 3      & 5      & 1   & 0.35400  & 5      & 5      & 1      & 5      & 2   & -0.54400 & 5      & 5      & 3      & 5      & 4   & 1.3800   \\
		1      & 3      & 1      & 3      & 2   & -0.48400 & 3      & 3      & 3      & 3      & 2   & -0.14900 & 5      & 5      & 5      & 5      & 4   & -0.15200 \\
		1      & 3      & 1      & 5      & 2   & -1.5870  & 3      & 3      & 3      & 5      & 2   & 0.84900  &   --    &    --    &   --     &    --    &   --  &    --     \\
		1      & 5      & 1      & 3      & 2   & -1.5870  & 3      & 5      & 3      & 3      & 2   & 0.84900  &     --   &  --    &     --   &   --     &  --   &   --     \\
		\hline  \hline 
	\end{tabular}
	\label{tab:sd_nn_effective_interaction}
\end{table}

We  implement Eq. (8.24) in Ref. \cite{suhonen2007nucleons} to transform the interaction matrix elements from the coupled-$JT$ basis to the selected SP basis. These transformed two-body interaction matrix elements are listed in Table \ref{tab:few_body_matrix_elements} (with indices from 12 to 631). It is noteworthy that an additional factor $(18/A)^{1/3}$ (with, e.g., $A=21$ for the $^{21}$O calculation) is necessary to scale the interaction matrix elements provided in Table \ref{tab:few_body_matrix_elements}. Indeed, these interaction matrix elements are also applicable to the spectral calculations of other oxygen isotopes using the corresponding scaling.

\centering
\renewcommand{\arraystretch}{0.8}
\setlength{\extrarowheight}{0.5pt}
\begin{longtable}{cccccc|cccccc|cccccc}
	\caption{The few-body Hamiltonian matrix elements (in MeV) in the SP basis. The indices $p$, $q$, $r$, and $s$ label the SP bases listed in Table \ref{tab:SP_basis}. The first 12 entries list the SP energies, while the rest give the two-body interaction matrix elements.
	}
	\label{tab:few_body_matrix_elements} \\
	\hline
	\hline
	Index & $\ \ p \ \ $ & $\ \ q\ \ $ & $\ \ s \ \ $ & $\ \ r\ \ $ & $\ \ \langle pq|H|rs\rangle \ \ $  &
	Index & $\ \ p \ \ $ & $\ \ q\ \ $ & $\ \ s \ \ $ & $\ \ r\ \ $ & $\ \ \langle pq|H|rs\rangle \ \ $  &
	Index & $\ \ p \ \ $ & $\ \ q\ \ $ & $\ \ s \ \ $ & $\ \ r\ \ $ & $\ \ \langle pq|H|rs\rangle \ \ $   \\
	\hline
	\endfirsthead
	\multicolumn{18}{c}{Table \thetable{} -- continued} \\
	\hline
	Index & $\ \ p \ \ $ & $\ \ q\ \ $ & $\ \ s \ \ $ & $\ \ r\ \ $ & $\ \ \langle pq|H|rs\rangle \ \ $  &
	Index & $\ \ p \ \ $ & $\ \ q\ \ $ & $\ \ s \ \ $ & $\ \ r\ \ $ & $\ \ \langle pq|H|rs\rangle \ \ $  &
	Index & $\ \ p \ \ $ & $\ \ q\ \ $ & $\ \ s \ \ $ & $\ \ r\ \ $ & $\ \ \langle pq|H|rs\rangle \ \ $   \\
	\hline
	\endhead
	0   & 0  & --  & 0 & --  &       -3.2079    & 211 & 0  & 10 & 8  & 5  & -0.061880 & 422 & 0 & 3  & 9  & 7 & -0.56058 \\
	1   & 1  & -- & 1 & --  &      -3.2079     & 212 & 0  & 10 & 11 & 7  & -0.37540  & 423 & 0 & 8  & 3  & 0 & 0.79350   \\
	2   & 2  & --  &  2 &  -- &     2.1117     & 213 & 0  & 10 & 10 & 8  & 0.23742 & 424 & 0 & 8  & 8  & 0 & 0.15667 \\
	3   & 3  & --  &  3 & --  &     2.1117     & 214 & 1  & 4  & 5  & 0  & -0.36893 & 425 & 0 & 8  & 2  & 1 & 0.45813 \\
	4   & 4  & --  & 4 &  -- &    2.1117      & 215 & 1  & 4  & 10 & 0  & 1.1222 & 426 & 0 & 8  & 7  & 1 & 0.84051 \\
	5   & 5  & --  &  5 &  -- &     2.1117     & 216 & 1  & 4  & 4  & 1  & -0.27100   & 427 & 0 & 8  & 4  & 2 & -0.64317 \\
	6   & 6  & --  & 6 & --  &     -3.9257    & 217 & 1  & 4  & 9  & 1  & -0.79350  & 428 & 0 & 8  & 9  & 2 & -0.28859 \\
	7   & 7  & --  & 7 &  -- &     -3.9257     & 218 & 1  & 4  & 11 & 2  & 0.27551 & 429 & 0 & 8  & 8  & 3 & 0.089222 \\
	8   & 8  & --  & 8 & --  &    -3.9257      & 219 & 1  & 4  & 5  & 3  & 0.011258 & 430 & 0 & 8  & 7  & 4 & 0.12136 \\
	9   & 9  & --  & 9 &  -- &    -3.9257      & 220 & 1  & 4  & 10 & 3  & 0.041559 & 431 & 0 & 8  & 6  & 5 & -0.046120 \\
	10  & 10 & -- & 10 & --  &    -3.9257      & 221 & 1  & 4  & 9  & 4  & -0.20968 & 432 & 0 & 8  & 10 & 6 & -0.26544 \\
	11  & 11 & -- & 11 &  -- &      -3.9257     & 222 & 1  & 4  & 8  & 5  & 0.22515 & 433 & 0 & 8  & 9  & 7 & 0.16788 \\
	12  & 0  & 1  & 1       & 0 & -1.9100    & 223 & 1  & 4  & 11 & 7  & 0.88636 & 434 & 1 & 2  & 3  & 0 & -0.36893 \\
	13  & 0  & 1  & 5       & 2 & -0.19375 & 224 & 1  & 4  & 10 & 8  & -0.56058 & 435 & 1 & 2  & 8  & 0 & 0.45813 \\
	14  & 0  & 1  & 4       & 3 & 0.193747 & 225 & 1  & 9  & 5  & 0  & -0.45813 & 436 & 1 & 2  & 2  & 1 & 0.15500    \\
	15  & 0  & 1  & 11      & 6 & -0.90298 & 226 & 1  & 9  & 10 & 0  & 0.84051 & 437 & 1 & 2  & 7  & 1 & -0.64789 \\
	16  & 0  & 1  & 10      & 7 & 0.902976 & 227 & 1  & 9  & 4  & 1  & -0.79350  & 438 & 1 & 2  & 4  & 2 & 0.0065000   \\
	17  & 0  & 1  & 9       & 8 & -0.90298 & 228 & 1  & 9  & 9  & 1  & 0.15667 & 439 & 1 & 2  & 9  & 2 & 0.15090 \\
	18  & 0  & 4  & 4       & 0 & -0.058000   & 229 & 1  & 9  & 11 & 2  & 0.046122 & 440 & 1 & 2  & 8  & 3 & -0.15728 \\
	19  & 0  & 4  & 9       & 0 & 0.79350   & 230 & 1  & 9  & 5  & 3  & 0.64317 & 441 & 1 & 2  & 7  & 4 & 0.075223 \\
	20  & 0  & 4  & 3       & 1 & -0.42600   & 231 & 1  & 9  & 10 & 3  & -0.12136 & 442 & 1 & 2  & 6  & 5 & 0.092929 \\
	21  & 0  & 4  & 8       & 1 & -0.79350  & 232 & 1  & 9  & 9  & 4  & -0.089220 & 443 & 1 & 2  & 10 & 6 & 0.51174 \\
	22  & 0  & 4  & 5       & 2 & 0.0065000   & 233 & 1  & 9  & 8  & 5  & 0.28859 & 444 & 1 & 2  & 9  & 7 & -0.32365 \\
	23  & 0  & 4  & 10      & 2 & 0.195194 & 234 & 1  & 9  & 11 & 7  & 0.26544 & 445 & 1 & 7  & 3  & 0 & -1.1222 \\
	24  & 0  & 4  & 4       & 3 & 0.0065000   & 235 & 1  & 9  & 10 & 8  & -0.16788 & 446 & 1 & 7  & 8  & 0 & 0.84051 \\
	25  & 0  & 4  & 9       & 3 & -0.058770 & 236 & 2  & 11 & 5  & 0  & 0.092929 & 447 & 1 & 7  & 2  & 1 & -0.64789 \\
	26  & 0  & 4  & 8       & 4 & -0.12152 & 237 & 2  & 11 & 10 & 0  & -0.24771 & 448 & 1 & 7  & 7  & 1 & -0.43767 \\
	27  & 0  & 4  & 7       & 5 & 0.24642 & 238 & 2  & 11 & 4  & 1  & 0.27551 & 449 & 1 & 7  & 4  & 2 & 0.90958 \\
	28  & 0  & 4  & 11      & 6 & 0.66065 & 239 & 2  & 11 & 9  & 1  & 0.046122 & 450 & 1 & 7  & 9  & 2 & 0.061880  \\
	29  & 0  & 4  & 10      & 7 & 0.132131 & 240 & 2  & 11 & 11 & 2  & 0.25986 & 451 & 1 & 7  & 8  & 3 & -0.19281 \\
	30  & 0  & 4  & 9       & 8 & -0.52852 & 241 & 2  & 11 & 5  & 3  & -0.50737 & 452 & 1 & 7  & 7  & 4 & 0.15820 \\
	31  & 0  & 9  & 4       & 0 & 0.79350   & 242 & 2  & 11 & 10 & 3  & -0.098150 & 453 & 1 & 7  & 6  & 5 & 0.24771 \\
	32  & 0  & 9  & 9       & 0 & -0.14050  & 243 & 2  & 11 & 9  & 4  & -0.091980 & 454 & 1 & 7  & 10 & 6 & 0.37540 \\
	33  & 0  & 9  & 3       & 1 & 0.79350   & 244 & 2  & 11 & 8  & 5  & -0.21659 & 455 & 1 & 7  & 9  & 7 & -0.23742 \\
	34  & 0  & 9  & 8       & 1 & 0.89150   & 245 & 2  & 11 & 11 & 7  & -0.23444 & 456 & 2 & 4  & 3  & 0 & 0.011258 \\
	35  & 0  & 9  & 5       & 2 & -0.55700   & 246 & 2  & 11 & 10 & 8  & -0.069930 & 457 & 2 & 4  & 8  & 0 & -0.64317 \\
	36  & 0  & 9  & 10      & 2 & -0.29271 & 247 & 3  & 5  & 5  & 0  & 0.0065000   & 458 & 2 & 4  & 2  & 1 & 0.0065000   \\
	37  & 0  & 9  & 4       & 3 & -0.55700   & 248 & 3  & 5  & 10 & 0  & -0.90958 & 459 & 2 & 4  & 7  & 1 & 0.90958 \\
	38  & 0  & 9  & 9       & 3 & -0.021860 & 249 & 3  & 5  & 4  & 1  & 0.011258 & 460 & 2 & 4  & 4  & 2 & -0.14900   \\
	39  & 0  & 9  & 8       & 4 & 0.16662 & 250 & 3  & 5  & 9  & 1  & 0.64317 & 461 & 2 & 4  & 9  & 2 & -0.48134 \\
	40  & 0  & 9  & 7       & 5 & -0.061880 & 251 & 3  & 5  & 11 & 2  & -0.50737 & 462 & 2 & 4  & 8  & 3 & 0.46317 \\
	41  & 0  & 9  & 11      & 6 & -0.29678 & 252 & 3  & 5  & 5  & 3  & -0.14900   & 463 & 2 & 4  & 7  & 4 & -0.13100   \\
	42  & 0  & 9  & 10      & 7 & -0.059360 & 253 & 3  & 5  & 10 & 3  & -0.13100   & 464 & 2 & 4  & 6  & 5 & -0.50737 \\
	43  & 0  & 9  & 9       & 8 & 0.23742 & 254 & 3  & 5  & 9  & 4  & 0.46317 & 465 & 2 & 4  & 10 & 6 & -0.46568 \\
	44  & 1  & 3  & 4       & 0 & -0.42600   & 255 & 3  & 5  & 8  & 5  & -0.48134 & 466 & 2 & 4  & 9  & 7 & 0.29452 \\
	45  & 1  & 3  & 9       & 0 & 0.79350   & 256 & 3  & 5  & 11 & 7  & -0.46568 & 467 & 2 & 9  & 3  & 0 & 0.22515 \\
	46  & 1  & 3  & 3       & 1 & -0.058000   & 257 & 3  & 5  & 10 & 8  & 0.29452 & 468 & 2 & 9  & 8  & 0 & -0.28859 \\
	47  & 1  & 3  & 8       & 1 & -0.79350  & 258 & 3  & 10 & 5  & 0  & 0.075223 & 469 & 2 & 9  & 2  & 1 & 0.15090 \\
	48  & 1  & 3  & 5       & 2 & 0.0065000   & 259 & 3  & 10 & 10 & 0  & -0.15820  & 470 & 2 & 9  & 7  & 1 & 0.061880  \\
	49  & 1  & 3  & 10      & 2 & 0.24642 & 260 & 3  & 10 & 4  & 1  & 0.041559 & 471 & 2 & 9  & 4  & 2 & -0.48134 \\
	50  & 1  & 3  & 4       & 3 & 0.0065000   & 261 & 3  & 10 & 9  & 1  & -0.12136 & 472 & 2 & 9  & 9  & 2 & 0.057421 \\
	51  & 1  & 3  & 9       & 3 & -0.12152 & 262 & 3  & 10 & 11 & 2  & -0.098150 & 473 & 2 & 9  & 8  & 3 & -0.42483 \\
	52  & 1  & 3  & 8       & 4 & -0.058770 & 263 & 3  & 10 & 5  & 3  & -0.13100   & 474 & 2 & 9  & 7  & 4 & -0.45895 \\
	53  & 1  & 3  & 7       & 5 & 0.19519 & 264 & 3  & 10 & 10 & 3  & -0.026540 & 475 & 2 & 9  & 6  & 5 & -0.21659 \\
	54  & 1  & 3  & 11      & 6 & 0.66065 & 265 & 3  & 10 & 9  & 4  & -0.62394 & 476 & 2 & 9  & 10 & 6 & -0.44060  \\
	55  & 1  & 3  & 10      & 7 & 0.13213 & 266 & 3  & 10 & 8  & 5  & -0.45895 & 477 & 2 & 9  & 9  & 7 & -0.41134 \\
	56  & 1  & 3  & 9       & 8 & -0.52852 & 267 & 3  & 10 & 11 & 7  & -0.41685 & 478 & 3 & 8  & 3  & 0 & -0.20968 \\
	57  & 1  & 8  & 4       & 0 & -0.79350  & 268 & 3  & 10 & 10 & 8  & -0.58144 & 479 & 3 & 8  & 8  & 0 & 0.089222 \\
	58  & 1  & 8  & 9       & 0 & 0.89150   & 269 & 4  & 9  & 5  & 0  & -0.15728 & 480 & 3 & 8  & 2  & 1 & -0.15728 \\
	59  & 1  & 8  & 3       & 1 & -0.79350  & 270 & 4  & 9  & 10 & 0  & 0.19281 & 481 & 3 & 8  & 7  & 1 & -0.19281 \\
	60  & 1  & 8  & 8       & 1 & -0.14050  & 271 & 4  & 9  & 4  & 1  & -0.20968 & 482 & 3 & 8  & 4  & 2 & 0.46317 \\
	61  & 1  & 8  & 5       & 2 & 0.55700    & 272 & 4  & 9  & 9  & 1  & -0.089220 & 483 & 3 & 8  & 9  & 2 & -0.42483 \\
	62  & 1  & 8  & 10      & 2 & 0.061880  & 273 & 4  & 9  & 11 & 2  & -0.091980 & 484 & 3 & 8  & 8  & 3 & -0.60574 \\
	63  & 1  & 8  & 4       & 3 & 0.55700    & 274 & 4  & 9  & 5  & 3  & 0.46317 & 485 & 3 & 8  & 7  & 4 & -0.62394 \\
	64  & 1  & 8  & 9       & 3 & -0.16662 & 275 & 4  & 9  & 10 & 3  & -0.62394 & 486 & 3 & 8  & 6  & 5 & -0.091980 \\
	65  & 1  & 8  & 8       & 4 & 0.021856 & 276 & 4  & 9  & 9  & 4  & -0.60574 & 487 & 3 & 8  & 10 & 6 & -0.41587 \\
	66  & 1  & 8  & 7       & 5 & 0.29271 & 277 & 4  & 9  & 8  & 5  & -0.42483 & 488 & 3 & 8  & 9  & 7 & -0.93210  \\
	67  & 1  & 8  & 11      & 6 & 0.29678 & 278 & 4  & 9  & 11 & 7  & -0.41587 & 489 & 4 & 7  & 3  & 0 & 0.041559 \\
	68  & 1  & 8  & 10      & 7 & 0.059355 & 279 & 4  & 9  & 10 & 8  & -0.93210  & 490 & 4 & 7  & 8  & 0 & 0.12136 \\
	69  & 1  & 8  & 9       & 8 & -0.23742 & 280 & 5  & 8  & 5  & 0  & 0.15090 & 491 & 4 & 7  & 2  & 1 & 0.075223 \\
	70  & 2  & 5  & 1       & 0 & -0.19375 & 281 & 5  & 8  & 10 & 0  & -0.061880 & 492 & 4 & 7  & 7  & 1 & 0.15820 \\
	71  & 2  & 5  & 4       & 0 & 0.0065000   & 282 & 5  & 8  & 4  & 1  & 0.22515 & 493 & 4 & 7  & 4  & 2 & -0.13100   \\
	72  & 2  & 5  & 9       & 0 & -0.55700   & 283 & 5  & 8  & 9  & 1  & 0.28859 & 494 & 4 & 7  & 9  & 2 & -0.45895 \\
	73  & 2  & 5  & 3       & 1 & 0.0065000   & 284 & 5  & 8  & 11 & 2  & -0.21659 & 495 & 4 & 7  & 8  & 3 & -0.62394 \\
	74  & 2  & 5  & 8       & 1 & 0.55700    & 285 & 5  & 8  & 5  & 3  & -0.48134 & 496 & 4 & 7  & 7  & 4 & -0.026540 \\
	75  & 2  & 5  & 5       & 2 & -0.90150  & 286 & 5  & 8  & 10 & 3  & -0.45895 & 497 & 4 & 7  & 6  & 5 & -0.098150 \\
	76  & 2  & 5  & 10      & 2 & -0.39301 & 287 & 5  & 8  & 9  & 4  & -0.42483 & 498 & 4 & 7  & 10 & 6 & -0.41685 \\
	77  & 2  & 5  & 4       & 3 & 0.75250   & 288 & 5  & 8  & 8  & 5  & 0.057421 & 499 & 4 & 7  & 9  & 7 & -0.58144 \\
	78  & 2  & 5  & 9       & 3 & 0.16045 & 289 & 5  & 8  & 11 & 7  & -0.44060  & 500 & 5 & 6  & 3  & 0 & 0.27551 \\
	79  & 2  & 5  & 8       & 4 & 0.16045 & 290 & 5  & 8  & 10 & 8  & -0.41134 & 501 & 5 & 6  & 8  & 0 & -0.046120 \\
	80  & 2  & 5  & 7       & 5 & -0.39301 & 291 & 7  & 11 & 5  & 0  & 0.51174 & 502 & 5 & 6  & 2  & 1 & 0.092929 \\
	81  & 2  & 5  & 11      & 6 & -1.6392 & 292 & 7  & 11 & 10 & 0  & -0.37540  & 503 & 5 & 6  & 7  & 1 & 0.24771 \\
	82  & 2  & 5  & 10      & 7 & 1.27853 & 293 & 7  & 11 & 4  & 1  & 0.88636 & 504 & 5 & 6  & 4  & 2 & -0.50737 \\
	83  & 2  & 5  & 9       & 8 & -1.0982 & 294 & 7  & 11 & 9  & 1  & 0.26544 & 505 & 5 & 6  & 9  & 2 & -0.21659 \\
	84  & 2  & 10 & 4       & 0 & 0.19519 & 295 & 7  & 11 & 11 & 2  & -0.23444 & 506 & 5 & 6  & 8  & 3 & -0.091980 \\
	85  & 2  & 10 & 9       & 0 & -0.29271 & 296 & 7  & 11 & 5  & 3  & -0.46568 & 507 & 5 & 6  & 7  & 4 & -0.098150 \\
	86  & 2  & 10 & 3       & 1 & 0.24642 & 297 & 7  & 11 & 10 & 3  & -0.41685 & 508 & 5 & 6  & 6  & 5 & 0.25986 \\
	87  & 2  & 10 & 8       & 1 & 0.061880  & 298 & 7  & 11 & 9  & 4  & -0.41587 & 509 & 5 & 6  & 10 & 6 & -0.23444 \\
	88  & 2  & 10 & 5       & 2 & -0.39301 & 299 & 7  & 11 & 8  & 5  & -0.44060  & 510 & 5 & 6  & 9  & 7 & -0.069930 \\
	89  & 2  & 10 & 10      & 2 & 0.18383 & 300 & 7  & 11 & 11 & 7  & -0.84271 & 511 & 6 & 10 & 3  & 0 & 0.88636 \\
	90  & 2  & 10 & 4       & 3 & -0.39301 & 301 & 7  & 11 & 10 & 8  & 0.43685 & 512 & 6 & 10 & 8  & 0 & -0.26544 \\
	91  & 2  & 10 & 9       & 3 & -0.20642 & 302 & 8  & 10 & 5  & 0  & -0.32365 & 513 & 6 & 10 & 2  & 1 & 0.51174 \\
	92  & 2  & 10 & 8       & 4 & -0.29117 & 303 & 8  & 10 & 10 & 0  & 0.23742 & 514 & 6 & 10 & 7  & 1 & 0.37540 \\
	93  & 2  & 10 & 7       & 5 & -0.27397 & 304 & 8  & 10 & 4  & 1  & -0.56058 & 515 & 6 & 10 & 4  & 2 & -0.46568 \\
	94  & 2  & 10 & 11      & 6 & -0.23444 & 305 & 8  & 10 & 9  & 1  & -0.16788 & 516 & 6 & 10 & 9  & 2 & -0.44060  \\
	95  & 2  & 10 & 10      & 7 & -0.32289 & 306 & 8  & 10 & 11 & 2  & -0.069930 & 517 & 6 & 10 & 8  & 3 & -0.41587 \\
	96  & 2  & 10 & 9       & 8 & -0.088450 & 307 & 8  & 10 & 5  & 3  & 0.29452 & 518 & 6 & 10 & 7  & 4 & -0.41685 \\
	97  & 3  & 4  & 1       & 0 & 0.19375 & 308 & 8  & 10 & 10 & 3  & -0.58144 & 519 & 6 & 10 & 6  & 5 & -0.23444 \\
	98  & 3  & 4  & 4       & 0 & 0.006500   & 309 & 8  & 10 & 9  & 4  & -0.93210  & 520 & 6 & 10 & 10 & 6 & -0.84271 \\
	99  & 3  & 4  & 9       & 0 & -0.55700   & 310 & 8  & 10 & 8  & 5  & -0.41134 & 521 & 6 & 10 & 9  & 7 & 0.43685 \\
	100 & 3  & 4  & 3       & 1 & 0.0065000   & 311 & 8  & 10 & 11 & 7  & 0.43685 & 522 & 7 & 9  & 3  & 0 & -0.56058 \\
	101 & 3  & 4  & 8       & 1 & 0.55700    & 312 & 8  & 10 & 10 & 8  & -0.42829 & 523 & 7 & 9  & 8  & 0 & 0.16788 \\
	102 & 3  & 4  & 5       & 2 & 0.75250   & 313 & 0  & 11 & 11 & 0  & -0.73483 & 524 & 7 & 9  & 2  & 1 & -0.32365 \\
	103 & 3  & 4  & 10      & 2 & -0.39301 & 314 & 0  & 11 & 5  & 1  & 1.4487 & 525 & 7 & 9  & 7  & 1 & -0.23742 \\
	104 & 3  & 4  & 4       & 3 & -0.90150  & 315 & 0  & 11 & 10 & 1  & 0.66449 & 526 & 7 & 9  & 4  & 2 & 0.29452 \\
	105 & 3  & 4  & 9       & 3 & 0.16045 & 316 & 0  & 11 & 11 & 3  & -0.31980  & 527 & 7 & 9  & 9  & 2 & -0.41134 \\
	106 & 3  & 4  & 8       & 4 & 0.16045 & 317 & 0  & 11 & 5  & 4  & -1.0169 & 528 & 7 & 9  & 8  & 3 & -0.93210  \\
	107 & 3  & 4  & 7       & 5 & -0.39301 & 318 & 0  & 11 & 10 & 4  & 0.18068 & 529 & 7 & 9  & 7  & 4 & -0.58144 \\
	108 & 3  & 4  & 11      & 6 & 1.0381 & 319 & 0  & 11 & 9  & 5  & -0.046120 & 530 & 7 & 9  & 6  & 5 & -0.069930 \\
	109 & 3  & 4  & 10      & 7 & -1.3988 & 320 & 0  & 11 & 11 & 8  & -0.29678 & 531 & 7 & 9  & 10 & 6 & 0.43685 \\
	110 & 3  & 4  & 9       & 8 & 1.5791 & 321 & 0  & 11 & 10 & 9  & 0.39817 & 532 & 7 & 9  & 9  & 7 & -0.42829 \\
	111 & 3  & 9  & 4       & 0 & -0.058770 & 322 & 1  & 5  & 11 & 0  & 1.4487 & 533 & 0 & 2  & 2  & 0 & -0.48400   \\
	112 & 3  & 9  & 9       & 0 & -0.021860 & 323 & 1  & 5  & 5  & 1  & -0.48400   & 534 & 0 & 2  & 7  & 0 & 0.64789  \\
	113 & 3  & 9  & 3       & 1 & -0.12152 & 324 & 1  & 5  & 10 & 1  & -0.64789 & 535 & 0 & 2  & 6  & 1 & -1.4487 \\
	114 & 3  & 9  & 8       & 1 & -0.16662 & 325 & 1  & 5  & 11 & 3  & 0.32916 & 536 & 0 & 2  & 3  & 2 & 0.013000    \\
	115 & 3  & 9  & 5       & 2 & 0.16045 & 326 & 1  & 5  & 5  & 4  & 0.013000    & 537 & 0 & 2  & 8  & 2 & 0.18029 \\
	116 & 3  & 9  & 10      & 2 & -0.20642 & 327 & 1  & 5  & 10 & 4  & -0.29441 & 538 & 0 & 2  & 7  & 3 & -0.29441 \\
	117 & 3  & 9  & 4       & 3 & 0.16045 & 328 & 1  & 5  & 9  & 5  & 0.18029 & 539 & 0 & 2  & 6  & 4 & 0.32916 \\
	118 & 3  & 9  & 9       & 3 & -0.34133 & 329 & 1  & 5  & 11 & 8  & 0.72371 & 540 & 0 & 2  & 9  & 6 & 0.72371 \\
	119 & 3  & 9  & 8       & 4 & -0.76453 & 330 & 1  & 5  & 10 & 9  & -0.97096 & 541 & 0 & 2  & 8  & 7 & -0.97096 \\
	120 & 3  & 9  & 7       & 5 & -0.29117 & 331 & 1  & 10 & 11 & 0  & 0.66449 & 542 & 0 & 7  & 2  & 0 & 0.64789  \\
	121 & 3  & 9  & 11      & 6 & -0.18598 & 332 & 1  & 10 & 5  & 1  & -0.64789 & 543 & 0 & 7  & 7  & 0 & 0.45383 \\
	122 & 3  & 9  & 10      & 7 & -0.71326 & 333 & 1  & 10 & 10 & 1  & 0.45383 & 544 & 0 & 7  & 6  & 1 & 0.66449 \\
	123 & 3  & 9  & 9       & 8 & -0.52727 & 334 & 1  & 10 & 11 & 3  & -0.067700  & 545 & 0 & 7  & 3  & 2 & -0.45479 \\
	124 & 4  & 8  & 4       & 0 & -0.12152 & 335 & 1  & 10 & 5  & 4  & 0.45479 & 546 & 0 & 7  & 8  & 2 & -0.25146 \\
	125 & 4  & 8  & 9       & 0 & 0.16662 & 336 & 1  & 10 & 10 & 4  & -0.17504 & 547 & 0 & 7  & 7  & 3 & 0.17504 \\
	126 & 4  & 8  & 3       & 1 & -0.058770 & 337 & 1  & 10 & 9  & 5  & 0.25146 & 548 & 0 & 7  & 6  & 4 & 0.067700   \\
	127 & 4  & 8  & 8       & 1 & 0.021856 & 338 & 1  & 10 & 11 & 8  & 0.13272 & 549 & 0 & 7  & 9  & 6 & -0.13272 \\
	128 & 4  & 8  & 5       & 2 & 0.16045 & 339 & 1  & 10 & 10 & 9  & -0.17807 & 550 & 0 & 7  & 8  & 7 & 0.17807 \\
	129 & 4  & 8  & 10      & 2 & -0.29117 & 340 & 3  & 11 & 11 & 0  & -0.31980  & 551 & 1 & 6  & 2  & 0 & -1.4487 \\
	130 & 4  & 8  & 4       & 3 & 0.16045 & 341 & 3  & 11 & 5  & 1  & 0.32916 & 552 & 1 & 6  & 7  & 0 & 0.66449 \\
	131 & 4  & 8  & 9       & 3 & -0.76453 & 342 & 3  & 11 & 10 & 1  & -0.067700  & 553 & 1 & 6  & 6  & 1 & -0.73483 \\
	132 & 4  & 8  & 8       & 4 & -0.34133 & 343 & 3  & 11 & 11 & 3  & 0.13314 & 554 & 1 & 6  & 3  & 2 & 1.0169 \\
	133 & 4  & 8  & 7       & 5 & -0.20642 & 344 & 3  & 11 & 5  & 4  & -0.58587 & 555 & 1 & 6  & 8  & 2 & 0.046122 \\
	134 & 4  & 8  & 11      & 6 & -0.18598 & 345 & 3  & 11 & 10 & 4  & -0.26354 & 556 & 1 & 6  & 7  & 3 & -0.18068 \\
	135 & 4  & 8  & 10      & 7 & -0.71326 & 346 & 3  & 11 & 9  & 5  & -0.46713 & 557 & 1 & 6  & 6  & 4 & 0.31980 \\
	136 & 4  & 8  & 9       & 8 & -0.52727 & 347 & 3  & 11 & 11 & 8  & -0.47311 & 558 & 1 & 6  & 9  & 6 & 0.29678 \\
	137 & 5  & 7  & 4       & 0 & 0.24642 & 348 & 3  & 11 & 10 & 9  & -0.12112 & 559 & 1 & 6  & 8  & 7 & -0.39817 \\
	138 & 5  & 7  & 9       & 0 & -0.061880 & 349 & 4  & 5  & 11 & 0  & -1.0169 & 560 & 2 & 3  & 2  & 0 & 0.013000    \\
	139 & 5  & 7  & 3       & 1 & 0.19519 & 350 & 4  & 5  & 5  & 1  & 0.013000    & 561 & 2 & 3  & 7  & 0 & -0.45479 \\
	140 & 5  & 7  & 8       & 1 & 0.29271 & 351 & 4  & 5  & 10 & 1  & 0.45479 & 562 & 2 & 3  & 6  & 1 & 1.0169 \\
	141 & 5  & 7  & 5       & 2 & -0.39301 & 352 & 4  & 5  & 11 & 3  & -0.58587 & 563 & 2 & 3  & 3  & 2 & -0.14900   \\
	142 & 5  & 7  & 10      & 2 & -0.27397 & 353 & 4  & 5  & 5  & 4  & -0.14900   & 564 & 2 & 3  & 8  & 2 & -0.32089 \\
	143 & 5  & 7  & 4       & 3 & -0.39301 & 354 & 4  & 5  & 10 & 4  & 0.52401 & 565 & 2 & 3  & 7  & 3 & 0.52401 \\
	144 & 5  & 7  & 9       & 3 & -0.29117 & 355 & 4  & 5  & 9  & 5  & -0.32089 & 566 & 2 & 3  & 6  & 4 & -0.58587 \\
	145 & 5  & 7  & 8       & 4 & -0.20642 & 356 & 4  & 5  & 11 & 8  & -0.32929 & 567 & 2 & 3  & 9  & 6 & -0.32929 \\
	146 & 5  & 7  & 7       & 5 & 0.18383 & 357 & 4  & 5  & 10 & 9  & 0.44178 & 568 & 2 & 3  & 8  & 7 & 0.44178 \\
	147 & 5  & 7  & 11      & 6 & -0.23444 & 358 & 4  & 10 & 11 & 0  & 0.18068 & 569 & 2 & 8  & 2  & 0 & 0.18029 \\
	148 & 5  & 7  & 10      & 7 & -0.32289 & 359 & 4  & 10 & 5  & 1  & -0.29441 & 570 & 2 & 8  & 7  & 0 & -0.25146 \\
	149 & 5  & 7  & 9       & 8 & -0.088450 & 360 & 4  & 10 & 10 & 1  & -0.17504 & 571 & 2 & 8  & 6  & 1 & 0.046122 \\
	150 & 6  & 11 & 1       & 0 & -0.90298 & 361 & 4  & 10 & 11 & 3  & -0.26354 & 572 & 2 & 8  & 3  & 2 & -0.32089 \\
	151 & 6  & 11 & 4       & 0 & 0.66065 & 362 & 4  & 10 & 5  & 4  & 0.52401 & 573 & 2 & 8  & 8  & 2 & -0.18786 \\
	152 & 6  & 11 & 9       & 0 & -0.29678 & 363 & 4  & 10 & 10 & 4  & -0.61429 & 574 & 2 & 8  & 7  & 3 & -0.70650  \\
	153 & 6  & 11 & 3       & 1 & 0.66065 & 364 & 4  & 10 & 9  & 5  & -0.70650  & 575 & 2 & 8  & 6  & 4 & -0.46713 \\
	154 & 6  & 11 & 8       & 1 & 0.29678 & 365 & 4  & 10 & 11 & 8  & -0.71062 & 576 & 2 & 8  & 9  & 6 & -0.72200   \\
	155 & 6  & 11 & 5       & 2 & -1.6392 & 366 & 4  & 10 & 10 & 9  & -0.73674 & 577 & 2 & 8  & 8  & 7 & -0.41134 \\
	156 & 6  & 11 & 10      & 2 & -0.23444 & 367 & 5  & 9  & 11 & 0  & -0.046120 & 578 & 3 & 7  & 2  & 0 & -0.29441 \\
	157 & 6  & 11 & 4       & 3 & 1.0381 & 368 & 5  & 9  & 5  & 1  & 0.18029 & 579 & 3 & 7  & 7  & 0 & 0.17504 \\
	158 & 6  & 11 & 9       & 3 & -0.18598 & 369 & 5  & 9  & 10 & 1  & 0.25146 & 580 & 3 & 7  & 6  & 1 & -0.18068 \\
	159 & 6  & 11 & 8       & 4 & -0.18598 & 370 & 5  & 9  & 11 & 3  & -0.46713 & 581 & 3 & 7  & 3  & 2 & 0.52401 \\
	160 & 6  & 11 & 7       & 5 & -0.23444 & 371 & 5  & 9  & 5  & 4  & -0.32089 & 582 & 3 & 7  & 8  & 2 & -0.70650  \\
	161 & 6  & 11 & 11      & 6 & -1.4899 & 372 & 5  & 9  & 10 & 4  & -0.70650  & 583 & 3 & 7  & 7  & 3 & -0.61429 \\
	162 & 6  & 11 & 10      & 7 & 0.64721 & 373 & 5  & 9  & 9  & 5  & -0.18786 & 584 & 3 & 7  & 6  & 4 & -0.26354 \\
	163 & 6  & 11 & 9       & 8 & -0.30186 & 374 & 5  & 9  & 11 & 8  & -0.72200   & 585 & 3 & 7  & 9  & 6 & -0.71062 \\
	164 & 7  & 10 & 1       & 0 & 0.90300 & 375 & 5  & 9  & 10 & 9  & -0.41134 & 586 & 3 & 7  & 8  & 7 & -0.73674 \\
	165 & 7  & 10 & 4       & 0 & 0.132131 & 376 & 8  & 11 & 11 & 0  & -0.29678 & 587 & 4 & 6  & 2  & 0 & 0.32916 \\
	166 & 7  & 10 & 9       & 0 & -0.059360 & 377 & 8  & 11 & 5  & 1  & 0.72371 & 588 & 4 & 6  & 7  & 0 & 0.067700   \\
	167 & 7  & 10 & 3       & 1 & 0.13213 & 378 & 8  & 11 & 10 & 1  & 0.13272 & 589 & 4 & 6  & 6  & 1 & 0.31980 \\
	168 & 7  & 10 & 8       & 1 & 0.059355 & 379 & 8  & 11 & 11 & 3  & -0.47311 & 590 & 4 & 6  & 3  & 2 & -0.58587 \\
	169 & 7  & 10 & 5       & 2 & 1.2785 & 380 & 8  & 11 & 5  & 4  & -0.32929 & 591 & 4 & 6  & 8  & 2 & -0.46713 \\
	170 & 7  & 10 & 10      & 2 & -0.32289 & 381 & 8  & 11 & 10 & 4  & -0.71062 & 592 & 4 & 6  & 7  & 3 & -0.26354 \\
	171 & 7  & 10 & 4       & 3 & -1.3988 & 382 & 8  & 11 & 9  & 5  & -0.72200   & 593 & 4 & 6  & 6  & 4 & 0.13314 \\
	172 & 7  & 10 & 9       & 3 & -0.71326 & 383 & 8  & 11 & 11 & 8  & -0.49736 & 594 & 4 & 6  & 9  & 6 & -0.47311 \\
	173 & 7  & 10 & 8       & 4 & -0.71326 & 384 & 8  & 11 & 10 & 9  & 0.46335 & 595 & 4 & 6  & 8  & 7 & -0.12112 \\
	174 & 7  & 10 & 7       & 5 & -0.32289 & 385 & 9  & 10 & 11 & 0  & 0.398167 & 596 & 6 & 9  & 2  & 0 & 0.72371 \\
	175 & 7  & 10 & 11      & 6 & 0.64721 & 386 & 9  & 10 & 5  & 1  & -0.97096 & 597 & 6 & 9  & 7  & 0 & -0.13272 \\
	176 & 7  & 10 & 10      & 7 & -0.93736 & 387 & 9  & 10 & 10 & 1  & -0.17807 & 598 & 6 & 9  & 6  & 1 & 0.29678 \\
	177 & 7  & 10 & 9       & 8 & 0.85443 & 388 & 9  & 10 & 11 & 3  & -0.12112 & 599 & 6 & 9  & 3  & 2 & -0.32929 \\
	178 & 8  & 9  & 1       & 0 & -0.90298 & 389 & 9  & 10 & 5  & 4  & 0.441783 & 600 & 6 & 9  & 8  & 2 & -0.72200   \\
	179 & 8  & 9  & 4       & 0 & -0.52852 & 390 & 9  & 10 & 10 & 4  & -0.73674 & 601 & 6 & 9  & 7  & 3 & -0.71062 \\
	180 & 8  & 9  & 9       & 0 & 0.23742 & 391 & 9  & 10 & 9  & 5  & -0.41134 & 602 & 6 & 9  & 6  & 4 & -0.47311 \\
	181 & 8  & 9  & 3       & 1 & -0.52852 & 392 & 9  & 10 & 11 & 8  & 0.46335 & 603 & 6 & 9  & 9  & 6 & -0.49736 \\
	182 & 8  & 9  & 8       & 1 & -0.23742 & 393 & 9  & 10 & 10 & 9  & -0.77364 & 604 & 6 & 9  & 8  & 7 & 0.46335 \\
	183 & 8  & 9  & 5       & 2 & -1.0982 & 394 & 1  & 11 & 11 & 1  & 0.75100    & 605 & 7 & 8  & 2  & 0 & -0.97096 \\
	184 & 8  & 9  & 10      & 2 & -0.088450 & 395 & 1  & 11 & 11 & 4  & -0.23559 & 606 & 7 & 8  & 7  & 0 & 0.17807 \\
	185 & 8  & 9  & 4       & 3 & 1.5791 & 396 & 1  & 11 & 10 & 5  & 0.18249 & 607 & 7 & 8  & 6  & 1 & -0.39817 \\
	186 & 8  & 9  & 9       & 3 & -0.52727 & 397 & 4  & 11 & 11 & 1  & -0.23559 & 608 & 7 & 8  & 3  & 2 & 0.44178 \\
	187 & 8  & 9  & 8       & 4 & -0.52727 & 398 & 4  & 11 & 11 & 4  & -0.16150  & 609 & 7 & 8  & 8  & 2 & -0.41134 \\
	188 & 8  & 9  & 7       & 5 & -0.088450 & 399 & 4  & 11 & 10 & 5  & -0.88885 & 610 & 7 & 8  & 7  & 3 & -0.73674 \\
	189 & 8  & 9  & 11      & 6 & -0.30186 & 400 & 4  & 11 & 11 & 9  & -0.84507 & 611 & 7 & 8  & 6  & 4 & -0.12112 \\
	190 & 8  & 9  & 10      & 7 & 0.85443 & 401 & 5  & 10 & 11 & 1  & 0.18249 & 612 & 7 & 8  & 9  & 6 & 0.46335 \\
	191 & 8  & 9  & 9       & 8 & -1.2827 & 402 & 5  & 10 & 11 & 4  & -0.88885 & 613 & 7 & 8  & 8  & 7 & -0.77364 \\
	192 & 0  & 5  & 5       & 0 & 0.15500    & 403 & 5  & 10 & 10 & 5  & -0.62050  & 614 & 0 & 6  & 6  & 0 & 0.75100    \\
	193 & 0  & 5  & 10      & 0 & 0.64789  & 404 & 5  & 10 & 11 & 9  & -1.0910 & 615 & 0 & 6  & 7  & 2 & -0.18249 \\
	194 & 0  & 5  & 4       & 1 & -0.36893 & 405 & 9  & 11 & 11 & 4  & -0.84507 & 616 & 0 & 6  & 6  & 3 & 0.23559  \\
	195 & 0  & 5  & 9       & 1 & -0.45813 & 406 & 9  & 11 & 10 & 5  & -1.0910 & 617 & 2 & 7  & 6  & 0 & -0.18249 \\
	196 & 0  & 5  & 11      & 2 & 0.092929 & 407 & 9  & 11 & 11 & 9  & -0.15200   & 618 & 2 & 7  & 7  & 2 & -0.62050  \\
	197 & 0  & 5  & 5       & 3 & 0.0065000   & 408 & 5  & 11 & 11 & 5  & -1.3090   & 619 & 2 & 7  & 6  & 3 & -0.88885 \\
	198 & 0  & 5  & 10      & 3 & 0.075223 & 409 & 5  & 11 & 11 & 10 & -1.3800    & 620 & 2 & 7  & 8  & 6 & -1.0910 \\
	199 & 0  & 5  & 9       & 4 & -0.15728 & 410 & 10 & 11 & 11 & 5  & -1.3800    & 621 & 3 & 6  & 6  & 0 & 0.23559  \\
	200 & 0  & 5  & 8       & 5 & 0.15090 & 411 & 10 & 11 & 11 & 10 & -0.15200   & 622 & 3 & 6  & 7  & 2 & -0.88885 \\
	201 & 0  & 5  & 11      & 7 & 0.51174 & 412 & 0  & 3  & 3  & 0  & -0.27100   & 623 & 3 & 6  & 6  & 3 & -0.16150  \\
	202 & 0  & 5  & 10      & 8 & -0.32365 & 413 & 0  & 3  & 8  & 0  & 0.79350   & 624 & 3 & 6  & 8  & 6 & -0.84507 \\
	203 & 0  & 10 & 5       & 0 & 0.64789  & 414 & 0  & 3  & 2  & 1  & -0.36893 & 625 & 6 & 8  & 7  & 2 & -1.0910 \\
	204 & 0  & 10 & 10      & 0 & -0.43767 & 415 & 0  & 3  & 7  & 1  & -1.1222 & 626 & 6 & 8  & 6  & 3 & -0.84507 \\
	205 & 0  & 10 & 4       & 1 & 1.1222 & 416 & 0  & 3  & 4  & 2  & 0.011258 & 627 & 6 & 8  & 8  & 6 & -0.15200   \\
	206 & 0  & 10 & 9       & 1 & 0.84051 & 417 & 0  & 3  & 9  & 2  & 0.22515 & 628 & 2 & 6  & 6  & 2 & -1.3090   \\
	207 & 0  & 10 & 11      & 2 & -0.24771 & 418 & 0  & 3  & 8  & 3  & -0.20968 & 629 & 2 & 6  & 7  & 6 & -1.3800    \\
	208 & 0  & 10 & 5       & 3 & -0.90958 & 419 & 0  & 3  & 7  & 4  & 0.041559 & 630 & 6 & 7  & 6  & 2 & -1.3800    \\
	209 & 0  & 10 & 10      & 3 & -0.15820  & 420 & 0  & 3  & 6  & 5  & 0.27551 & 631 & 6 & 7  & 7  & 6 & -0.15200   \\
	210 & 0  & 10 & 9       & 4 & 0.19281 & 421 & 0  & 3  & 10 & 6  & 0.88636 &  --   &  -- &  --  &  --  &  -- &  --         \\ 
	\hline \hline
\end{longtable}

\end{document}